\documentclass[journal=jacsat,manuscript=article]{achemso}
\usepackage[version=3]{mhchem} 
\usepackage{amsmath}
\usepackage{graphicx}

\author{Heiko Dumlich}
\affiliation{Fachbereich Physik, Freie Universit\"{a}t Berlin, 14195 Berlin, Germany}
\author{John Robertson}
\affiliation{Department of Engineering, University of Cambridge, Cambridge CB2 1PZ, United Kingdom}
\author{Stephanie Reich}
\email{reich@physik.fu-berlin.de}
\affiliation{Fachbereich Physik, Freie Universit\"{a}t Berlin, 14195 Berlin, Germany}

\title{Nanotube caps on Ni, Fe, and NiFe nano particles: A path to chirality selective growth}

\begin{document}


\begin{abstract}
Carbon nanotubes have properties depending on the arrangement of carbon atoms on the tube walls, called chirality. Also 
it has been tried to grow nanotubes of only one chirality for more than a decade it is still not possible today. A narrowing 
of the distribution of chiralities, however, which is a first step towards chirality control, has been observed for the 
growth of nanotubes on catalysts composed of nickel and iron. In this paper, we have calculated carbon-metal bond energies, 
adhesion energies and charge distributions of carbon nanotube caps on Ni, Fe and NiFe alloy clusters using density functional 
theory. A growth model using the calculated energies was able to reproduce the experimental data of the nanotube growth on 
the alloy catalysts. The electronic charge was found to be redistributed from the catalyst particles to the edges of the 
nanotube caps in dependence of the chiral angles of the caps increasing the reactivity of the edge atoms. Our study develops 
an explanation for the chirality enrichment in the carbon nanotube growth on alloy catalyst particles.
\end{abstract}



\section{Introduction}
Carbon nanotubes have mechanical and electrical properties, depending on the helicity/atomic structure of the carbon atoms 
on their tubes surface (chirality).~\cite{S.Reich2004,RiichiroSaito1992,TeriWangOdom1998} The self assembling growth process 
of nanotubes generates nanotubes with various chiralities in the same growth ensemble.\cite{SergeiM.Bachilo2002} Three phases 
have been suggested to account for the chirality distribution found in nanotube ensembles, the nucleation 
phase,~\cite{StephanieReich2006} the growth phase,~\cite{FengDing2009,Dumlich2010} and the termination of the 
growth.~\cite{Boerjesson2011} The nucleation phase involves the formation of graphene-like sheets (sheets of hexagonally 
arranged carbon atoms) which need to have pentagons in their structure to induce curvature.~\cite{Ohta2009} Only with the 
curvature, which is achieved by a curved template particle, it is possible for a graphene sheet to transform into a carbon 
nanotube cap and lift off the particle.~\cite{Schebarchov2011,DiegoA.Gomez-Gualdron2012} This cap can then elongate with a 
chirality dependent growth rate by carbon addition to the edge of the growing tube.~\cite{FengDing2009,Dumlich2010,RahulRao2012} 
Many theoretical and experimental studies tried to understand how the self assembly of nanotubes works, however, it is still 
not fully understood how the chirality can be controlled during the growth of carbon 
nanotubes.~\cite{SumioIijima1992,J.Gavillet2001,Jean-YvesRaty2005,StephanieReich2006,Yazyev2008,FengDing2009,Ohta2009,Qiang2010,ErikC.Neyts2010,RahulRao2012,Fouquet2012} 

In recent experimental studies it was found that growth on alloy NiFe particles leads to an enrichment of certain 
chiralities.~\cite{Wei-HungChiang2009,Wei-HungChiang2009a} A theoretical study on a nickel particle suggested that the electronic 
charge transfer might be important to control the chirality-selective growth process.~\cite{QiangWang2011} Another theoretical 
study attempted to explain the chirality enrichment on the NiFe alloy particles by the lower excess energies for certain 
chiralities.~\cite{DebosrutiDutta2012} The study, however, considered plane surfaces and elongated nanotube caps (non-minimal seed 
caps with only an inferior number of growth sites/kinks), which does not seem to be appropriate to describe the chirality 
selection on a catalyst particle.~\cite{DebosrutiDutta2012,Dumlich2013PhD}

In this paper we study the $\left(5,5\right)$ armchair and $\left(9,0\right)$ zigzag carbon nanotube caps connected to various 
Ni, NiFe and Fe catalyst particles. We calculate the average carbon-metal bond energies, adhesion energies and charge 
distributions using density functional theory. We find the highest adhesion and lowest excess energy for armchair and zigzag 
caps on the $\text{Ni}_{27}\text{Fe}_{28}$ alloy cluster. Small energy differences between the armchair and zigzag bonds 
allow to derive carbon addition barriers. The barriers can be used in a growth model leading to chirality distributions that 
compare satisfactorily with the recent experimental results of growth on NiFe alloy particles. A charge transfer from the 
catalyst particles to the caps induces a dipole moment between the catalyst particle and the cap. The polarity of the bond 
between the cap edge and catalyst atoms increases with increasing Fe content. The charge redistribution is found to depend 
on the chirality, as the line density of edge sites increases with higher chiral angle. The excess electron charges on the 
armchair rim atoms are found to be between $\left(2.90\pm0.06\right)~\text{e}$ for Ni and $\left(4.15\pm0.14\right)~\text{e}$ 
for Fe. The increasing reactivity induced through the excess electron charges on the edge carbon cap atoms allows to explain 
why the nanotube growth rate on iron is higher than on nickel.

\section{Methodology}
\subsection{Computational Methods}
We performed spin polarized density functional theory calculations with the \emph{ab-initio} package 
SIESTA.~\cite{PabloOrdejon1996,JoseSoler2002} We used the generalized gradient approximation parameterized by Perdew, Burke 
and Ernzerhof,~\cite{JohnP.Perdew1996} as the bias towards compact cluster structures is reduced compared to the local 
density approximation.~\cite{Baletto2005} The calculations used norm conserving nonlocal pseudopotentials generated in the 
scheme of Troullier and Martins with the parameters presented in the PhD thesis of Dumlich.~\cite{Troullier1991,Dumlich2013PhD} 
The valence electrons were described by localized pseudoatomic orbitals. To balance the computational time and the accuracy 
to a reasonable level, we used a double-$\zeta$ polarized (DZP) basis set. The cutoff radii of the orbitals were chosen with 
$r_s=6.099~\text{Bohr}$ and $r_p=7.832~\text{Bohr}$ for the $s$ and $p$ orbital of the carbon atoms, $r_s=9.649~\text{Bohr}$ 
and $r_d=6.001~\text{Bohr}$ for the $s$ and $d$ orbital of the iron atoms, and $r_s=9.187~\text{Bohr}$ and 
$r_d=5.572~\text{Bohr}$ for the $s$ and $d$ orbital of the nickel atoms. The mesh-cutoff for the real-space integration 
corresponded to about $350~\text{Ry}$. We used only the $\Gamma$-point to calculate the total energies, as all studied 
systems have finite dimensions.

For our calculations we consider the situation, in which a carbon nanotube cap has already formed on a catalyst particle, but 
is not elongated. The systems studied in this paper therefore consist of two parts, a catalytic particle and a carbon nanotube 
cap. We consider the $\text{Fe}_{55}$, $\text{Ni}_{12}\text{Fe}_{43}$, $\text{Ni}_{27}\text{Fe}_{28}$, and $\text{Ni}_{55}$ 
clusters to understand the influence of alloy systems and chiralities on the cap-cluster interaction.

\begin{figure}
\begin{center}\includegraphics[scale=0.20]{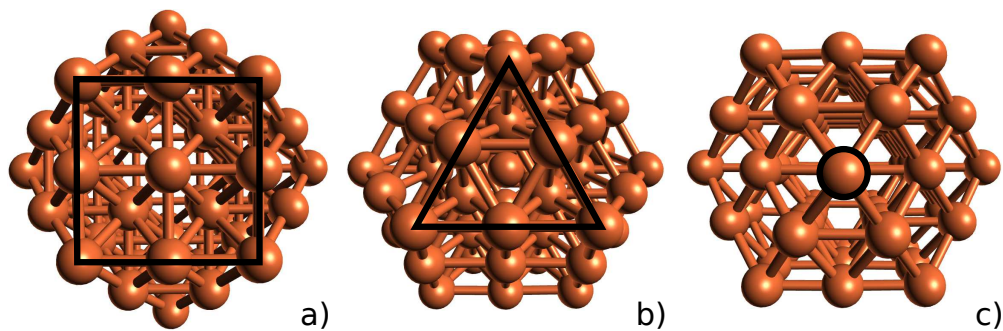}\end{center}
\caption{\label{fig:fig1}Top view on a ball and stick sketch model of the spots of a $55$ atom (iron) cluster. The atoms that form the top of the cluster are marked. a) Spot 1, fcc(100) with a $3$ times $3$ atom \emph{square} ($9$ atoms). b) Spot 2, fcc(111) with a $3$ atom sided \emph{triangle} ($6$ atoms). c) Spot 3, with only $1$ atom at the \emph{top} of the cluster.}
\end{figure}

Our (deformed) icosahedral catalyst particles consists of 55 atoms~\cite{Singh2008} that initially 
form a higly symmetric structure containing six fcc(100), eight fcc(111) surfaces, and three 
distinctive spots to add a carbon nanotube cap, see \ref{fig:fig1} a) to \ref{fig:fig1} 
c). The icosahedral particles have been found to be the most stable configuration.~\cite{Singh2008} The 
relaxed particle structures do not keep the icosahedral symmetry and are not expected to represent the 
global minimum, as it is not feasible to find the global minimum of the clusters in density functional 
theory.~\cite{Zhu2010,Baletto2005} The energy differences between different geometry optimized catalyst 
clusters range between $2~\text{eV}$ and $3~\text{eV}$, corresponding to previously reported 
values.~\cite{Zhu2010} We calculated the energies of the geometry optimized catalyst structures from each 
individual initial structure of the combined system to derive consistent carbon-metal and adhesion 
energies.

The carbon nanotube caps were created with the program code CaGe using the isolated pentagon rule (IPR), 
which states that caps are energetically most stable, if all six pentagons needed for the cap inclination 
are isolated from each other.~\cite{G.Brinkmann2010,G.Brinkmann1999,S.Reich2005} We created and geometry 
optimized fullerene structures starting from the as-generated caps of CaGe to get to a decent cap structure. 
There are many cap configurations for a certain chirality, which hinders to draw conclusions for the most 
chiralities.~\cite{G.Brinkmann1999,S.Reich2005} The isolated pentagon rule used for the generation of our caps, 
however, allows to reduce the number of possible caps, by only considering energetically favorable 
structures.~\cite{S.Reich2005} Especially the $\left(5,5\right)$ ($30$ atom cap) and $\left(9,0\right)$ 
($39$ atom cap) chiralities only have one possible cap structure which fulfills the isolated pentagon 
rule,~\cite{L.Li2006} making them perfect candidates for our study.

\begin{figure}
\begin{center}\includegraphics[scale=0.20]{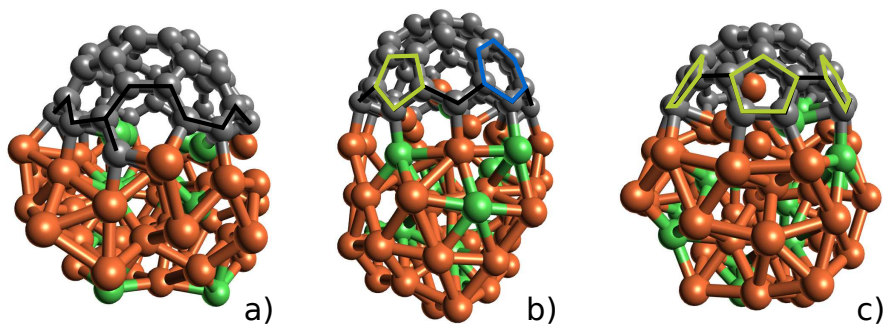}\end{center}
\caption{\label{fig:fig2} Ball and sticks sketch model of carbon nanotube caps on a $\mbox{Ni}_{12}\mbox{Fe}_{43}$ alloy cluster on three different spots. Carbon atoms in grey, nickel atoms in green, and iron atoms in orange. (a) Spot 1: $\left(9,0\right)$ cap shows a Klein-edge through bond break of a pentagon at the rim. (b) Spot 2: $\left(9,0\right)$ cap showing a hexagon and pentagon at the zigzag rim. (c) Spot 3: $\left(5,5\right)$ cap showing an armchair rim formed exclusively by pentagons.}
\end{figure}

The caps were transferred on the three different spots of the catalyst particles, see \ref{fig:fig1}. The fit of the 
rims of the carbon nanotube caps to the spots of the clusters were performed by hand. In \ref{fig:fig2} we show 
geometry optimized structures of the $\left(9,0\right)$ and $\left(5,5\right)$ cap bound to the three spots of the 
$\text{Ni}_{12}\text{Fe}_{43}$ catalyst as examples for all the systems we studied. The atoms at the rim of the cap bind 
to the cluster, see \ref{fig:fig2}. The deformation (structure change) of the cluster is significant for the 
presented alloy systems. The atoms in the rim of the cap also adjusted their positions, which suggests a dynamic process 
of carbon and metal reshaping.~\cite{StephanHofmann2007,DiegoA.Gomez-Gualdron2011}. The edges/rims of the caps are 
composed of only armchair and zigzag sites, which is true for all nanotubes. If a pentagon bond gets broken at the edge, 
a Klein-edge can form and offer a site for addition of a single carbon atom to close the edge with a hexagon, see 
\ref{fig:fig2} a). The Klein-edge configuration occurs commonly for the $\left(9,0\right)$ cap on the \emph{square} 
spot and for the $\left(5,5\right)$ cap on the \emph{top} spot and might be a possible way to avoid the initiation barrier 
for a new layer. The configuration with the Klein edge was recently suggested to be energetically favorable for the armchair 
growth of graphene (A5' site).~\cite{Artyukhov2012} In general there is a nearly continuous number of ways to combine the 
cap with the cluster in dependence of the chirality of the cap, however, the chosen spots are expected to allow for the best 
comparison.

All optimized geometries were relaxed to a maximal atomic force of $0.04~\text{eV/\AA}$. The total energy of the combined 
system as well as the energy of the cluster and cap were calculated in various structures and basis sets to account for 
the basis set superposition error (counterpoise correction).~\cite{Boys1970,Liu1973,Liu1989,Duijneveldt1994} The basis set 
superposition errors for our calculations range between $2.0~\text{eV}$ and $3.0~\text{eV}$ with an average error of 
$\left(2.5\pm0.3\right)~\text{eV}$. The adhesion energy between the carbon nanotube cap and the catalyst cluster can be 
calculated with:
\begin{equation}
E_{ad}=E_{tot}-E_{cap}^{relax}-E_{cluster}^{relax}-E^{BSSE},\label{eq:ECMNiFecapcluster}
\end{equation}
where $E_{tot}$ is the energy of the geometry optimized combined system of the cap and the catalyst cluster, $E_{cap}^{relax}$ 
is the energy of the geometry optimized cap, and $E_{cluster}^{relax}$ is the energy of the geometry optimized 
cluster.~\cite{Zhu2010,Wang2012}

To assess the stability of the combined system we compare the cap-on cluster system with other systems that contain the 
same number of carbon and metal atoms. This leads to formulas for the excess energy:
\begin{equation}
E_{x}^{i}=E_{tot}-E_{cluster}^{relax}-n_{\text{C}}E^{i}_{\text{C}}-E^{BSSE},\label{eq:Eexcess}
\end{equation}
with $E^{i}_{\text{C}}$ the energy per carbon atom for the system the combined cap-on-cluster system is compared to and 
$n_{\text{C}}$ the number of carbon atoms in the cap of the combined system.~\cite{X.Fan2003,StephanieReich2006,Zhu2010} We 
calculate the excess energy per atom in comparison to the energy of a system of an isolated metal cluster and a fullerene, to 
determine which system configuration is lower in energy and therefore more stable.~\cite{X.Fan2003,StephanieReich2006,Zhu2010} 
Alternatively the excess energy compared to the fullerene structure can be obtained by removal of the dangling bond contributions 
of the adhesion energy. This leads to
\begin{equation}
E_{x}^{\text{fullerene}}=E_{CM}=E_{ad}-2\cdot m\cdot E_a^{vac}-(n-m)\cdot E_z^{vac},\label{eq:Ecm} 
\end{equation}
where $E_{ad}$ is the adhesion energy, $E_a^{vac}$ is the armchair bond energy in vacuum, and $E_z^{vac}$ is the zigzag bond 
energy in vaccum. The factor $2\cdot m$ results from the number of armchair bonds at the rim of the nanotube cap and the 
factor $(n-m)$ results from the number of zigzag bonds at the rim of the cap.~\cite{Dumlich2010,Dumlich2010a} The vacuum bond 
energies (for a straight cut rim) can be derived with the equations 
$E_a^{vac}=\frac{E_{\text{C}_{60}}/2-E^{\left(5,5\right)}_{cap}}{2\cdot m}$ and 
$E_z^{vac}=\frac{E_{\text{C}_{78}}/2-E^{\left(9,0\right)}_{cap}}{n-m}$ where $E_{\text{C}_{i}}$ is the energy of a fullerene 
formed from two $\left(n,m\right)$ caps. We used the system specific vacuum bond energies to determine the carbon-metal bond 
energies. To determine the carbon-metal bond energies per bond we divided through the number of bonds at the edge $n+m$.

An error for the energies was estimated from the standard deviation by averaging over identical systems with small changes in their 
initial configuration. The standard deviations are rather large, as only two values were included which does not have a statistical 
significance, however, it allows to estimate the order of the error with at least $0.1~\text{eV}$ for the excess energy. The error 
for the adhesion energy is estimated to be slightly higher with about $0.7~\text{eV}$.

We performed Bader population analysis calculations on the cap-cluster systems to determine the electron charge transfer between the 
carbon cap atoms and the catalyst particle atoms.~\cite{Mulliken1955,Henkelman2006,Sanville2007,Tang2009}

\subsection{Growth model}
The growth model applied in this paper is based on the idea that the carbon addition occurs to the hexagonal rim of 
the nanotube.~\cite{Dumlich2010,Dumlich2010a} The hexagonal rim contains armchair and zigzag edges.~\cite{Dumlich2010,Dumlich2010a} 
The addition of carbon dimers elongates the rim with hexagons and transforms the rim through a change of the number of 
energetically favorable addition sites.~\cite{Dumlich2010,Dumlich2010a} An energy barrier occurs for the addition of carbon 
atoms to the rim, which is especially high for the addition to zigzag rims.~\cite{Dumlich2010,FengDing2009} The energy barrier 
for the rim transformation occuring through the carbon dimer addition to an armchair rim (initiation/closing of a new layer) 
is determined by
\begin{equation}
\Delta_{a}=2\cdot\left|E_{z}-E_{a}\right|,\label{eq:deltaac}
\end{equation}
where $E_a$ is the energy of an armchair carbon-metal bond and $E_z$ is the energy of a zigzag carbon-metal 
bond.~\cite{Dumlich2010,Dumlich2013PhD} The energy barriers influence the chirality dependent growth 
rate.~\cite{FengDing2009,Dumlich2010,RahulRao2012} A factor for the different growth rates in dependence of the chirality is 
given with
\begin{equation}
\Gamma\left(n,m\right)=\begin{cases}
\frac{\Lambda_{aa.aa}\left(n,m\right)\cdot\delta_a+\Lambda_{aa.z}\left(n,m\right)\cdot\delta_{az}}{n+m} & \text{if }2m-n>0,\\
\frac{\Lambda_{aa.z}\left(n,m\right)\cdot\delta_{az}}{n+m} & \text{otherwise},\end{cases}\label{eq:Gammaext}
\end{equation}
where $\delta_a=\exp\left(-\Delta_a/k_bT\right)$ is an exponential factor to account for the temperature dependence 
of the addition barrier to armchair sites and $\delta_{az}=\exp\left(-\Delta_{az}/k_bT\right)=1$, as an addition barrier 
for $aa.z$ sites (kinks) is negligible ($\Delta_{az}=0$).~\cite{Dumlich2010,Dumlich2013PhD,FengDing2009} The other terms 
are defined as $\Lambda_{aa.aa}=2m-n-1+1/\left(2m-n\right)$ and $\Lambda_{aa.z}=\min\left(m,n-m\right)$.~\cite{Dumlich2010} 
The growth rate factor $\Gamma\left(n,m\right)$ can have values between $0$ and $0.5$ in dependence of the chirality.

$\Gamma\left(n,m\right)$ allows to determine the relative difference in growth rate between different chiralities. To derive 
a chirality distribution we include the influence of the nucleation phase, i.e., whether a particular tube cap is 
nucleated or not.~\cite{S.Reich2005} We assume the tube diameters (and also the chirality) to be fixed by the 
nucleation.~\cite{S.Reich2006,StephanieReich2006} To consider the dependence of nanotube diameters on the diameter of the 
catalyst particles,~\cite{YimingLi2001,Inoue2005,Fiawoo2012} we therefore multiply Equation~(\ref{eq:Gammaext}) by a 
Gaussian distribution of the nanotube diameters $f\left(d;\mu,\sigma^2\right)$ and obtain a growth rate factor 
$\Gamma^*\left(n,m\right)$ which can be compared to a chirality distribution,
\begin{equation}
\Gamma^*\left(n,m\right)=\frac{1}{\sigma\sqrt{2\pi}}e^{-\frac{\left(d-\mu\right)^2}{2\sigma^2}}\cdot\Gamma\left(n,m\right).\label{eq:Gammastar}
\end{equation}
The tube diameter distribution might also include additional effects that do not result from the particle diameters and 
which might not be covered by the Gaussian distribution, however, the distribution serves the simplicity of the model.

\section{Results}
\subsection{Adhesion energies and carbon-metal bond energies of nanotube caps}
The adhesion energies are high with slightly more than $-20~\text{eV}$, varying about a few eV in dependence of the spot, catalyst and 
cap. Therefore the nanotube caps are not expected to lift off the catalyst particle spontaneously and perform dome 
closure.~\cite{YoungHeeLee1997,D.-H.Oh1998} The armchair caps have adhesion energies that are a few eV higher than the zigzag energies, 
which is a result of the number of dangling bonds at the edge of the caps. The zigzag caps have $n+m=9+0=9$ dangling bonds. The armchair 
caps have $n+m=5+5=10$ dangling bonds. Therefore the adhesion energy of the armchair cap is higher than the adhesion energy of the zigzag 
cap, even though the energy per dangling bond is lower for the armchair compared to the zigzag bond.~\cite{D.-H.Oh1998}

\begin{table}
\caption{\label{tab:averageadhesionexcessenergiesNiFe}Comparison of adhesion energies and carbon-metal bond energies of nanotube caps on metallic/alloy clusters in eV. The errors are standard deviations from averaging over the spots.}
\begin{center}
\begin{tabular}{ccccc}
$\text{cap}$ & $\text{particle}$ & $E_{ad}~\left(\text{eV}\right)$ & $E_{CM}/\text{bond}~\left(\text{eV}\right)$ & $E_{CM}~\left(\text{eV}\right)$ \\
\hline
$\left(5,5\right)$ & $\text{Fe}_{55}$               & $\left(-21.2\pm0.0\right)$ & $\left(0.43\pm0.00\right)$ & $4.3$ \\
$\left(5,5\right)$ & $\text{Ni}_{12}\text{Fe}_{43}$ & $\left(-23.2\pm1.1\right)$ & $\left(0.30\pm0.11\right)$ & $3.0$ \\
$\left(5,5\right)$ & $\text{Ni}_{27}\text{Fe}_{28}$ & $\left(-23.5\pm1.0\right)$ & $\left(0.27\pm0.07\right)$ & $2.7$ \\	
$\left(5,5\right)$ & $\text{Ni}_{55}$               & $\left(-23.2\pm0.3\right)$ & $\left(0.32\pm0.03\right)$ & $3.2$ \\
\\
$\left(9,0\right)$ & $\text{Fe}_{55}$               & $\left(-20.7\pm\text{-}.\text{-}\right)$ & $\left(0.40\pm\text{-}.\text{-}\right)$ & $3.6$ \\
$\left(9,0\right)$ & $\text{Ni}_{12}\text{Fe}_{43}$ & $\left(-21.2\pm0.9\right)$ & $\left(0.33\pm0.10\right)$ & $3.0$ \\
$\left(9,0\right)$ & $\text{Ni}_{27}\text{Fe}_{28}$ & $\left(-21.5\pm1.1\right)$ & $\left(0.30\pm0.13\right)$ & $2.7$ \\
$\left(9,0\right)$ & $\text{Ni}_{55}$               & $\left(-20.2\pm0.0\right)$ & $\left(0.46\pm0.00\right)$ & $4.1$ \\
\hline
\end{tabular}
\end{center}
\end{table}

Averaging the adhesion energies for the alloy compositions we find 
$E_{ad}^{\text{Fe}_{55}}=\left(-21.0\pm0.3\right)~\text{eV}$, $E_{ad}^{\text{Ni}_{12}\text{Fe}_{43}}=\left(-22.3\pm1.4\right)~\text{eV}$, 
$E_{ad}^{\text{Ni}_{27}\text{Fe}_{28}}=\left(-22.3\pm1.5\right)~\text{eV}$, and 
$E_{ad}^{\text{Ni}_{55}}=\left(-22.4\pm1.5\right)~\text{eV}$. Two effects account for the adhesion energy. One is the structure 
and the other is the material. The mixture of two materials distorts the catalyst structure as they have different electronic structure 
resulting in different bond lengths and lattice constants. We observe an increase of adhesion energy for the NiFe alloy systems compared 
to the pure Fe clusters, see \ref{tab:averageadhesionexcessenergiesNiFe}, the effect, however, is not significant.

The adhesion energies averaged on the three spots introduced in \ref{fig:fig1}, show no obvious trend either, 
with $E_{ad}^{1}=\left(-22.0\pm1.5\right)~\text{eV}$ for the \emph{square} spot 1, 
$E_{ad}^{2}=\left(-23.0\pm1.4\right)~\text{eV}$ for the \emph{triangle} spot 2, and 
$E_{ad}^{3}=\left(-21.7\pm1.0\right)~\text{eV}$ for the \emph{top} spot 3. We use this result and average over 
different spots when calculating adhesion energies as a function of the cap and alloy composition, see 
\ref{tab:averageadhesionexcessenergiesNiFe}. The \emph{triangle} spot shows the highest adhesion energy, but 
the adhesion energies calculated for the other spots are within the standard deviation of spot 2.

The excess/carbon-metal bond energies decrease non monotonically with increasing Ni content from Fe to Ni. The Ni 
cluster shows the highest excess energy for the $\left(9,0\right)$ cap. The lowest excess energy is observed for 
the $\text{Ni}_{27}\text{Fe}_{28}$, which also showed the highest adhesion energy, suggesting that the 
$\text{Ni}_{27}\text{Fe}_{28}$ alloy cluster yields the best growth conditions for the systems compared in this 
study. The low excess energy allows for a fast formation of caps on the $\text{Ni}_{27}\text{Fe}_{28}$ cluster 
and the higher adhesion energy prevents the cap lift off, which makes the NiFe alloy systems more stable compared 
to the pure elemental catalyst clusters.

\subsection{Charge distribution on cap and cluster atoms}
\begin{table}
\caption{\label{tab:chargedistributionNiFe}Charge redistributions between the carbon nanotube caps and the metallic/alloy clusters. $q_{\text{C}}^{\text{rim}}$ is the total charge shift of the carbon atoms at the rim of the nanotube cap, $q_{\text{C}}$ is the total charge shift considering all carbon atoms of the cap, $q_{\text{Fe}}$ is the total charge shift considering all iron atoms, and $q_{\text{Ni}}$ is the total charge shift considering all nickel atoms of the catalyst particle.}
\begin{center}
\begin{tabular}{cccccc}
$\text{cap}$ & $\text{particle}$ & $q_{\text{C}}^{\text{rim}}~\left(\text{e}\right)$ & $q_{\text{C}}~\left(\text{e}\right)$ & $q_{\text{Fe}}~\left(\text{e}\right)$ & $q_{\text{Ni}}~\left(\text{e}\right)$ \\
\hline
$\left(5,5\right)$ & $\text{Fe}_{55}$               & $\left(4.15\pm0.14\right)$ & $\left(5.31\pm0.72\right)$ & $\left(-5.32\pm0.72\right)$ & \\
$\left(5,5\right)$ & $\text{Ni}_{12}\text{Fe}_{43}$ & $\left(4.08\pm0.31\right)$ & $\left(5.06\pm0.52\right)$ & $\left(-7.18\pm0.65\right)$ & $\left(2.12\pm0.46\right)$  \\
$\left(5,5\right)$ & $\text{Ni}_{27}\text{Fe}_{28}$ & $\left(3.67\pm0.24\right)$ & $\left(4.36\pm0.36\right)$ & $\left(-6.46\pm0.20\right)$ & $\left(2.10\pm0.32\right)$  \\
$\left(5,5\right)$ & $\text{Ni}_{55}$               & $\left(2.90\pm0.06\right)$ & $\left(3.32\pm0.36\right)$ & 			    & $\left(-3.33\pm0.38\right)$ \\
\\
$\left(9,0\right)$ & $\text{Fe}_{55}$               & $\left(3.26\pm\text{-}.\text{-}\right)$ & $\left(4.68\pm\text{-}.\text{-}\right)$ & $\left(-4.68\pm\text{-}.\text{-}\right)$ & \\
$\left(9,0\right)$ & $\text{Ni}_{12}\text{Fe}_{43}$ & $\left(3.39\pm0.17\right)$ & $\left(4.62\pm0.22\right)$ & $\left(-6.71\pm0.42\right)$ & $\left(2.09\pm0.59\right)$  \\
$\left(9,0\right)$ & $\text{Ni}_{27}\text{Fe}_{28}$ & $\left(2.99\pm0.05\right)$ & $\left(3.96\pm0.02\right)$ & $\left(-6.48\pm0.29\right)$ & $\left(2.52\pm0.28\right)$  \\
$\left(9,0\right)$ & $\text{Ni}_{55}$               & $\left(2.34\pm0.08\right)$ & $\left(3.11\pm0.42\right)$ & 			    & $\left(-3.11\pm0.42\right)$ \\
\hline
\end{tabular}
\end{center}
\end{table}

The charge population on the atoms in the rim of the cap and for the metal atoms in the catalyst 
particle shows a stronger trend than the energies, see \ref{tab:chargedistributionNiFe}. All charge values 
are excess charges compared to the isolated atom valence electron configuration with 8 charges on each Fe atom, 
10 charges on each Ni atom and 4 charges on each C atom. The metal atoms partly loose their electrons to the 
carbon atoms in the cap, with whom they form carbon-metal bonds. The amount of electron charge transfer to 
the carbon atoms depends on the catalyst element and on the bond type of the edge atom (zigzag or armchair). 
The charge on the rim of the armchair cap is higher than on the rim of the zigzag cap. The trend weakens, 
but does not vanish, if the charge transfer per bond is considered, as the number of carbon-metal bonds is 
$10$ for the armchair and only $9$ for the zigzag cap. Considering the charge transfer to the whole cap and 
dividing through the number of bonds leads to an equal charge transfer to the armchair and zigzag caps per 
bond. The highest layer of the catalyst atoms supplies the major part of the electrons to the carbon nanotube 
cap. The carbon atoms of the cap that are not part of the rim have an average valence charge of about 4 e, 
with low deviations (below $0.1~\text{e}$), which means that they do not take part in the charge 
redistribution/polar binding between the metal catalyst and the carbon cap. The electron charge redistribution 
is localised at the outer rim atoms of the cap that form the carbon-metal bonds, corresponding to a polar bond. 
The localisation is slightly higher for the armchair edges with about $80\%$ of the charge localised at the 
outer rim atoms compared to about $74\%$ at the outer edge zigzag cap atoms, which is the reason why the 
armchair rim atoms yield a higher charge per bond.

The charge on the cap increases with Fe content and becomes maximal for the elemental Fe cluster. We observe 
the same behaviour at the rim, with the exclusion of the $\left(9,0\right)$ cap on Fe, which, however, may be 
an artifact, as only one cap was considered on spot 3. The higher charge points to a higher reactivity on Fe 
compared to Ni, which likely leads to a faster growth rate on iron compared to nickel, previously observed 
experimentally.~\cite{Yuan2011}

The Fe atoms do not only supply their electron charge to the C atoms, but also to the Ni atoms, see the charge 
transfer for the NiFe alloy systems in \ref{tab:chargedistributionNiFe}. The iron atoms loose about $5\text{-}7$ 
electron charges. The nickel atoms either gain about $2\text{-}2.5~\text{e}$ in the alloy systems or loose 
about $3~\text{e}$ if Ni is the only element in the catalyst particle. The carbon atoms in the cap always 
receive electron charges, see \ref{tab:chargedistributionNiFe}. The total amount of charge received by the 
carbon cap is about $3\text{-}5~\text{e}$. The amount of charge supplied to the rim is slightly lower with 
about $2\text{-}4~\text{e}$.

Following from the reactivity argument, which results from the increased charge on the edge atoms, we can give 
a geometric argument for the preference of armchair over zigzag structures in the following. It follows from 
the number of edge sites. We consider the line density of edge sites, which corresponds to the number of edges 
sites divided by the circumference of the nanotube rim
\begin{equation}
\lambda=\frac{N_a+N_z}{\left|\vec{C}_h\right|}=\frac{n+m}{a_0\cdot\sqrt{n^2+nm+m^2}},\label{eq:lambdaedgesitedensity}
\end{equation}
with $N_a+N_z$ the number of armchair and zigzag sites and $\left|\vec{C}_h\right|$ the circumference of the tube. 
The righternmost equation follows for straight rim configurations. Equation (\ref{eq:lambdaedgesitedensity}) leads 
to a line density of $\lambda_a=2/\left(\sqrt{3}a_0\right)$ for $\left(n=m\right)$ armchair and $\lambda_z=1/a_0$ for 
$\left(n\text{ integer},~m=0\right)$ zigzag tubes, for all other tubes $\left(n\neq m\neq0\right)$ the value of the 
line density is between $\lambda_a$ and $\lambda_z$ $\left(\lambda_a>\lambda_c>\lambda_z\right)$. Considering the 
fact that a higher density of edge sites increases the number of carbon metal bonds, directly gives the argument why 
armchair tubes are prefered compared to zigzag tubes, as the number of electrons at the edge is increased, yielding 
a higher reactivity. The comparison of the $\left(5,5\right)$ cap to the $\left(9,0\right)$ cap presented here, serves 
as an example as they have nearly identical diameters, but a different number of edge sites. The increased number of 
sites with higher chiral angles leads to an enrichment of armchair/near-armchair tubes in nanotube samples from simple 
geometric reasoning. This geometric argument might also be translated to growth rate considerations of nanotubes. The 
edges of tubes with higher chiral angles contain more sites for carbon atoms to dock, independent of the exact addition 
mechanism, supporting arguments for faster growth of higher chiral angle tubes.

\begin{table}
\caption{\label{tab:electricdipolemomentsNiFe}Electric dipole moments $\Delta$ between the nanotube caps and metallic clusters in Debye. The last column of the table shows electric dipole moments for the armchair/zigzag caps averaged over the spots of a specific catalyst composition.}
\begin{center}
\begin{tabular}{cccccc}
$\text{cap}$ & $\text{particle}$ & $\Delta_{avg}$~$\left(\text{Debye}\right)$ \\
\hline
$\left(5,5\right)$ & $\text{Fe}_{55}$                & 12.4 \\ 
$\left(9,0\right)$ & $\text{Fe}_{55}$                & 14.7 \\
$\left(5,5\right)$ & $\text{Ni}_{12}\text{Fe}_{43}$  & 10.6 \\ 
$\left(9,0\right)$ & $\text{Ni}_{12}\text{Fe}_{43}$  & 12.1 \\
$\left(5,5\right)$ & $\text{Ni}_{27}\text{Fe}_{28}$  & 10.0 \\ 
$\left(9,0\right)$ & $\text{Ni}_{27}\text{Fe}_{28}$  & 11.4 \\
$\left(5,5\right)$ & $\text{Ni}_{55}$                &  8.6 \\ 
$\left(9,0\right)$ & $\text{Ni}_{55}$                &  9.7 \\
\hline
\end{tabular}
\end{center}
\end{table}

The charge transfer also induces a dipole moment in the nanotube cap and cluster system, see 
\ref{tab:electricdipolemomentsNiFe}. The highest electric dipole moment can be found with $14.7~\text{Debye}$ for the 
$\left(9,0\right)$ cap on spot 3 of the Fe cluster and the lowest electric dipole moment is $7.4~\text{Debye}$ for the 
$\left(5,5\right)$ cap on the Ni cluster. A decrease of the electric dipole moment is correlated with the Ni content in 
the composition of the catalyst particle, where higher Ni content leads to lower electric dipole moments. The zigzag 
caps have higher electric dipole moments compared to the armchair caps, which is likely a result of the weaker localisation 
of electron charge at the carbon-metal bond forming atoms for the zigzag caps. The difference of the average electric 
dipole moments of a certain catalyst composition between armchair and zigzag caps decreases with increasing Ni content, 
with the highest difference of $2.3~\text{Debye}$ on Fe and the lowest difference of $1.1~\text{Debye}$ on Ni. The electric 
dipole moments generate an electric field, which was suggested to increase the landing probability of carbon atoms on the 
catalyst particle.~\cite{Mohammad2012} Therefore the landing probability on Fe is higher than on Ni, leading to a higher 
growth rate on catalysts containing Fe. A higher landing probability might also become a problem, if the carbon atoms cannot 
be incorporated at the edge faster than the carbon precursors land on the particle, as this might lead to amorphous carbon 
encapsulation of the catalyst and prevent further growth. The higher landing probability for zigzag caps can lead to a 
reduced number of zigzag tubes, as zigzag edges are expected to have a slower growth rate, which makes it especially hard 
to fulfill the requirement of fast carbon incorporation to prevent catalyst encapsulation.

\section{Discussion}
In the following we want to discuss the results we can derive from the energies and charge distributions calculated in 
this paper and compare them to other studies. There are two important results, that can be derived from the carbon-metal 
bond energies. First, the energies are important for the nucleation phase, as lower excess (carbon-metal bond) energies 
point to a higher formation probability.~\cite{StephanieReich2006,FengDing2008} Second, the carbon addition barrier 
$\Delta_a$ can be derived from the carbon-metal bond energies for the growth phase, which was found to influence the 
chirality dependent growth rate.~\cite{Dumlich2010,FengDing2009,RahulRao2012} Both phases determine the chirality of the 
carbon nanotube ensemble which is grown during the nanotube synthesis. We determine the barrier energies $\Delta_a$ 
(Equation~(\ref{eq:deltaac})), by using $E_a=E_{CM}^{\left(5,5\right)}/\text{bond}$ and 
$E_z=E_{CM}^{\left(9,0\right)}/\text{bond}$ of \ref{tab:averageadhesionexcessenergiesNiFe}. The barriers result from small 
deviations between armchair and zigzag bond energies. Previous studies considered flat metallic surfaces as catalysts 
and tried to optimize the fit between the catalyst surface and the edge of the cap.~\cite{StephanieReich2006,DebosrutiDutta2012} 
On a curved particle, as in this paper, a perfect fit between the edge of the nanotube cap and the catalyst particle is 
not possible, which increases the carbon-metal bond energies. The bond energies for armchair edges from Reich~\emph{et al.}, 
derived for a flat Ni surface, range from $E_a=0.12~\text{eV}$ to $E_a=1.12~\text{eV}$, comparing well to our average 
value $E_a^{\text{Ni}_{55}}=\left(0.32\pm0.04\right)~\text{eV}$.~\cite{StephanieReich2006} The values for zigzag edges 
from Reich~\emph{et al.} range from $E_z=0.16~\text{eV}$ to $E_z=1.44~\text{eV}$, which are also comparable to our value 
$E_z^{\text{Ni}_{55}}=0.46~\text{eV}$.~\cite{StephanieReich2006} The caps connect to various spots on the catalyst clusters 
which increases the deviation of the bond energies and therefore renders the chiral selectivity by structural fit of the 
cap and the catalyst particle unlikely. The large standard deviations point to a general problem for \emph{ab-initio} 
studies of carbon nanotube growth. The quantitative reproducibility is rather weak and it should be desired to test various 
systems with slightly different configurations/parameters. The energy barriers for the studied catalyst compositions are 
calculated with $\Delta_a^{\text{Fe}_{55}}=0.06~\text{eV}$, $\Delta_a^{\text{Ni}_{12}\text{Fe}_{43}}=0.06~\text{eV}$, 
$\Delta_a^{\text{Ni}_{27}\text{Fe}_{28}}=0.06~\text{eV}$, and $\Delta_a^{\text{Ni}_{55}}=0.28~\text{eV}$. The barriers 
are equal for all iron containing catalyst compositions.

\begin{figure}
\includegraphics[scale=0.45]{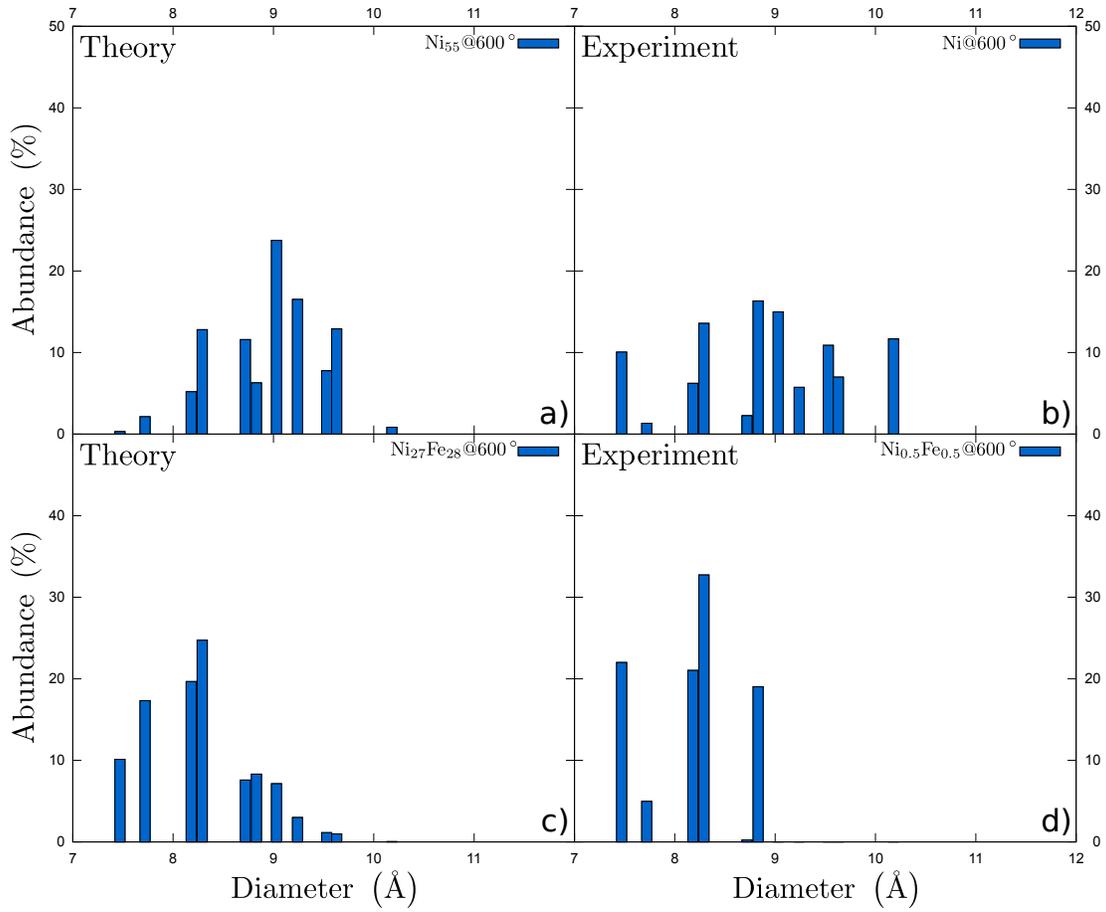}
\caption{\label{fig:fig3} Normalized abundances in dependence of the tube diameter; a) $\left(\mu=9.0~\text{\AA},\sigma=0.6~\text{\AA}\right)$ and c) $\left(\mu=8.1~\text{\AA},\sigma=0.6~\text{\AA}\right)$ estimated from our theoretical growth model, see Equation~(\ref{eq:Gammastar}); b) and d) from experimental photoluminescence data by Chiang~\emph{et al.} for nanotubes grown on Ni and on a nickel-iron alloy at $600^{\circ}$ C.~\cite{Wei-HungChiang2009}}
\end{figure}

The barriers can be inserted in Equation (\ref{eq:Gammastar}) to determine the chirality distributions, where 
the contribution of $\Gamma\left(n,m\right)$ (Equation (\ref{eq:Gammaext})) is equal for all systems that mainly 
contain iron atoms ($\delta_a=0.45$ for $600^{\circ}~\text{C}$) and only the gaussian diameter distribution factor 
changes the chirality distribution. We can compare the results of our model to the experimental results presented by 
Chiang~\emph{et al.}, who grew nanotubes on NiFe alloy systems to analyse the influence of the catalyst composition 
on the chirality distribution of a nanotube ensemble.~\cite{Wei-HungChiang2009,Wei-HungChiang2009a} They derived 
chirality distributions from photoluminescence data using calculated photoluminescence 
intensities.~\cite{Wei-HungChiang2009,Y.Oyama2006} The chirality distributions satisfactorily fit the results 
derived from our growth model, see \ref{fig:fig3}.~\cite{Wei-HungChiang2009} Especially the growth 
on iron containing systems is interesting, as the experimental study found the chirality distributions grown from 
catalyst with the composition $\text{Ni}_{0.27}\text{Fe}_{0.73}$ and $\text{Ni}_{0.5}\text{Fe}_{0.5}$ to be almost 
identical,~\cite{Wei-HungChiang2009} corresponding well to our results. The Ni catalyst particle has a significantly 
larger barrier energy, leading to a suppression of armchair growth sites, as, e.g. $\delta_a=0.024$ for 
$600^{\circ}~\text{C}$. In the experimental study the nanotubes grown on a Ni catalyst show a 
relatively wide chirality distribution with a peak for $\left(9,4\right)$.~\cite{Wei-HungChiang2009,Wei-HungChiang2009a} 
Especially important seems to be the diameter region of the nanotubes with $9.0~\text{\AA}$ for $\left(9,4\right)$ and 
$8.8~\text{\AA}$ for $\left(7,6\right)$ which have the highest intensity/abundance in the experimental 
study.~\cite{Wei-HungChiang2009} A slight descent of intensity occurs for chiralities with diameters that have 
smaller/higher tube diameters than about $9.0~\text{\AA}$, pointing to a lower number of catalyst particles, or other 
unknown effects, to grow tubes of that higher/lower diameters. Increasing the iron content of the composition of the 
catalyst particles until iron becomes the major component, leads to a significant narrowing of the chirality distribution 
to only a few chiralities at lower diameters, compared to the Ni catalyst particle, see \ref{fig:fig3} d). The fcc-lattice constant of iron $a_{\text{Fe}}=3.45~\text{\AA}$ is lower 
than the lattice constant of nickel $a_{\text{Ni}}=3.63~\text{\AA}$,~\cite{Dumlich2009} which leads to smaller diameter 
alloy catalyst particles with increasing Fe content, which might be a reason for the shift of the chirality distribution 
to lower diameter nanotubes ($\left(7,6\right)$ and $\left(8,4\right)$ abundance increased), as the diameter of the grown 
nanotubes depend on the diameter of the catalyst on which they are grown in the tangential growth mode/under growth 
conditions close to thermodynamic equilibrium.~\cite{YimingLi2001,Inoue2005,Fiawoo2012} The experimental study, however, 
tried to obtain equal particle diameters through the preparation process, pointing to a dependence on the material instead 
of the catalyst diameters.~\cite{Wei-HungChiang2009} Our model was not intended to perfectly reproduce all abundances, however, it 
still gives a fair approximation to the experimental results and successfully reproduces the most significant change in the 
chirality distribution by the change of the catalyst.

Besides the chirality distributions obtained from our growth model we also want to compare our charge transfer 
results to the literature. Wang~\emph{et al.} suggested that the short ranged charge distribution on nanotube edge 
atoms and catalyst atoms might be important for the chirality-selective growth of carbon nanotubes, as electron charges 
would increase the reactivity of the edge atoms.~\cite{QiangWang2011} We observe an increase of electron charge on the 
carbon edge atoms with charge supply by the metal atoms in agreement with Wang~\emph{et al.}.~\cite{QiangWang2011} The 
average charge values on the carbon rim atoms, see \ref{tab:chargedistributionNiFe}, compare well to the values 
calculated by Wang~\emph{et al.} for, e.g., the $\left(5,5\right)$ nanotube cap on nickel we find an average value of 
$0.29~\text{e}$ which compares to the slightly higher values of Wang~\emph{et al.} between $0.31~\text{e}$ and 
$0.38~\text{e}$. To determine the effect of the charges on chirality distributions, we also put a focus on alloy systems. 
In alloy systems a charge distribution between two metallic species leads to an electron accumulation not only on the 
carbon edge atoms, but also on the nickel atoms. We find higher charges on armchair than on zigzag edges, which was 
suggested by Wang~\emph{et al.} to be used to influence the chirality.~\cite{QiangWang2011} The alloy composition has a 
significant effect on the charge distribution. We find an increase of electron charge on the carbon cap edge atoms from 
Ni to Fe with increasing Fe content in the alloy, pointing to an increased growth rate of nanotubes through increased 
reactivity of the nanotube edge atoms, which compares well to the higher growth rates found for iron compared to 
nickel.~\cite{Yuan2011} Another relevant factor for the growth rate was found to be the metal d orbital 
energy.~\cite{Robertson2012} The charge distribution patterns suggested by Wang~\emph{et al.} resemble the edge 
structure of armchair and zigzag sites.

Theoretical studies can only model some aspects of the nanotube growth, neglecting other aspects, e.g., the 
effect of Ostwald ripening,~\cite{Boerjesson2011} that influence the chirality distribution as well. Further the 
chirality distributions determined in experiments have to be regarded with care as huge differences for the 
abundances of the chiralities might arise through the method used to determine the abundances, i.e. the intensity 
of a measured entity is not directly proportional to the abundance of the 
tube.~\cite{StephanieReich2005,Y.Oyama2006,SebastianHeeg2009} Therefore we did not expect to obtain results that 
perfectly match our growth model, however, we see it as a success that the model correctly describes the qualitative 
features of the chirality selective growth process, which suggests, that the model might include some part of the 
truth to solve the puzzle of chirality selective growth.

\section{Conclusions}
In summary we calculated adhesion energies, excess energies, and electronic charge redistributions between carbon 
nanotube caps and NiFe alloy systems using density functional theory. The highest adhesion energies and lowest 
excess energies are found for the $\text{Ni}_{27}\text{Fe}_{28}$ alloy cluster, for both armchair and zigzag caps. 
The energy differences between armchair and zigzag were found to be low. The curved form of the catalyst particle 
can be regarded as a constraint to the fit between the nanotube edge and the catalyst, which tends to lower the 
energy difference between armchair and zigzag caps. The small energy differences between the armchair and zigzag 
caps allow to derive carbon addition barriers, which - using the growth model presented in this paper - lead to 
chirality distributions that compare satisfactorily with experimental results. The charge transfer between the cap 
and the catalyst particles increases with increasing Fe content, which induces a dipole moment. The charge 
transfer to the armchair caps is higher than to the zigzag caps, in contrast to the electric dipole moment, which 
is higher for zigzag than for armchair caps and has a maximum of about $15~\text{Debye}$ on the iron particle. This 
is a consequence of the chirality dependent line density of edge sites, which decreases with lower chiral angle. The 
excess electron charges on the carbon rim atoms increase with Fe content of the catalyst particle from 
$\left(2.90\pm0.06\right)~\text{e}$ for Ni to $\left(4.15\pm0.14\right)~\text{e}$ for Fe. The excess electron charges 
increase the reactivity of the carbon cap atoms, which explains why the nanotube growth rate on iron is higher than 
on nickel. Our results will be useful for the understanding of the growth of carbon nanotubes on alloy catalysts.

\acknowledgement
We acknowledge S. Heeg and M. Fouquet for useful discussions. This work was supported by ERC under grant no. 210642.

\end{document}